\documentclass[USenglish,twocolumn]{article}

\ifx\directlua\undefined\ifx\XeTeXcharclass\undefined
  \usepackage[utf8]{inputenc}                           
  \else\RequirePackage[no-math]{fontspec}[2017/03/31]\fi 
  \else\RequirePackage[no-math]{fontspec}[2017/03/31]\fi 
\usepackage[big,online]{dgruyter}

\theoremstyle{dgthm}

\theoremstyle{dgdef}

\usepackage{graphicx} 
\graphicspath{{Figures/}{}}
\usepackage{amsmath}

\usepackage{amssymb}

\usepackage{siunitx}
\usepackage{physics}
\usepackage[style=ieee, doi=true,isbn=false,url=false, eprint=false,maxbibnames=99]{biblatex}
\AtEveryBibitem{\clearlist{language}}
\usepackage{hyperref}
\hypersetup{pdfstartview={FitH},pdfpagemode={UseNone},
            colorlinks,linkcolor=blue, citecolor=blue, urlcolor=blue,
            bookmarksopen=true, pdfnewwindow=true}
\usepackage[all]{hypcap}

\begin{document}

\articletype{Research Article}


\title{Inverse Design of Nonlinear Metasurfaces for Sum Frequency Generation}

\runningtitle{SFG Metasurface Optimization}
\author[1]{Neuton Li}
\author[1,2]{Jihua Zhang}
\author[1]{Dragomir N. Neshev}
\author[1]{Andrey A. Sukhorukov}
\affil[1]{ARC Centre of Excellence for Transformative Meta-Optical Systems (TMOS), Department of Electronic Materials Engineering, Research School of Physics, The Australian National University, Canberra, ACT~2600, Australia, neuton.li@anu.edu.au}
\affil[2]{Songshan Lake Materials Laboratory, Dongguan, Guangdong 523808, P. R. China}

\abstract{
Sum frequency generation (SFG) has multiple applications, from optical sources to imaging, where efficient conversion requires either long interaction distances or large field concentrations in a quadratic nonlinear material. Metasurfaces provide an essential avenue to enhanced SFG due to resonance with extreme field enhancements with an integrated ultrathin platform. In this work, we formulate a general theoretical framework for multi-objective topology optimization of nanopatterned metasurfaces that facilitate high-efficiency SFG and simultaneously select the emitted direction and tailor the metasurface polarization response. 
Based on this framework, we present novel metasurface designs showcasing ultimate flexibility in transforming the outgoing nonlinearly generated light for applications spanning from imaging to polarimetry. For example, one of our metasurfaces produces highly polarized and directional SFG emission with an efficiency of over \SI{0.2}{\centi\metre\squared\per\giga\watt} in a \SI{10}{\nano\metre} signal operating bandwidth.    
}

\maketitle

\section{Introduction}


Sum-frequency generation (SFG) is a fundamentally important second-order nonlinear process with many applications ranging from wavelength conversion of optical sources \cite{Doughty2022ConsiderationsImplementation} and infrared imaging \cite{Junaid2018Mid-infraredImaging, A.S.2019Mid-infraredPulses, Hanninen2018High-resolutionMicroscopy,Camacho-Morales2021InfraredMetasurfaces} to nonlinear polarimetry \cite{Zhu2022NonlinearUpconversion}. This phenomenon arises from induced polarizations in the medium, and in general, it can be observed in the presence of strong optical fields. Efficient second-order nonlinear frequency conversion, such as the SFG, 
traditionally requires long interaction lengths in bulky nonlinear crystals.
As a result, only certain crystalline orientations and input polarisations can satisfy the phase-matching conditions for producing sizable nonlinear effects. This limits the types of polarization transformations and the directionality of emission that are possible in nonlinear bulk crystals.  
   
Recent advances in nanotechnologies have facilitated the development of ultra-thin single-layer dielectric metasurfaces, where optical nano-resonators can enhance and tailor the nonlinear interactions with functionalities beyond the capabilities of traditional bulky crystals~\cite{Koshelev2023NonlinearPerspective,DeAngelis:2020:NonlinearMetaOptics, Grinblat2021NonlinearControl, Huang:2023-1:PRP, Kuznetsov:2024-roadmap:ACSP}. To generate optical resonances, previous metasurface designs have often relied on semi-analytical approaches in the limiting cases of Mie-type modes for individual nanoresonators~\cite{Marino2019Zero-OrderMetasurfaces, Kivshar2017Meta-OpticsResonances, Melik-Gaykazyan2018SelectiveNanoparticles, camacho2016nonlinear, Smirnova2018MultipolarNanoparticles, Frizyuk2019Second-harmonicMaterials, Camacho-Morales2016NonlinearNanoantennas}, or bound state in the continuum resonances~\cite{Vabishchevich2018EnhancedMetasurfaces, Carletti2019High-harmonicContinuum, Anthur2020ContinuousMetasurfaces, Fang:2022-2100498:LPR, Zograf2022High-HarmonicContinuum, Zhang:2022-2200031:LPR, Zheng:2022-125411:PRB, Kolkowski2023NonlinearMetasurfaces}. 
The angular-dependent properties of nonlocal metasurfaces could also be utilized to tune the nonlinear interactions over a range of wavelengths~\cite{Jiang2024TunableRange}. There is ongoing research on the enhancement of SFG in metasurfaces with resonances at non-degenerate wavelengths~\cite{Liu2018AnMixer, Yuan2019SecondMaterial, Camacho-Morales2022Sum-FrequencyModes, Rocco2022SecondPlatforms}. Non-planar structures with broken 3D symmetries have been identified for designing effective nonlinear susceptibility response~\cite{Rocco2020VerticalNanoantennas, Gigli2021TensorialMeta-optics}. Furthermore, several studies have considered inverse-design or machine learning approaches to optimize for strong resonances \cite{Xu2020EnhancedApproach, Sitawarin:2018-B82:PRJ, Hughes2018AdjointDevices, Raju2022MaximizedMetamaterials, Mann2023InverseGeneration}.
The resulting geometries have highly counterintuitive nanostructured geometries, which prove superior to conventional designs in their respective applications. 

However, these examples fall short of controlling nonlinear generation beyond the mere enhancement of conversion efficiency, which is only a part of the advantages that metasurfaces offer. For many practical applications such as imaging, it is often important to consider the input and output polarizations as well as the directional distribution of the generated emission. Control of the nonlinear polarization and the regulation of diffraction orders while maintaining high overall efficiency still remain a major challenge in the field. High-efficiency nonlinear conversion generally requires strong field enhancements in the nonlinear region, which can be characterized by the Q-factor of the resonance, its matching with the pulse bandwidth, and the overlap between high Q-factor modes. The task of engineering overlapping high Q-factor resonances with a prescribed bandwidth in materials at non-degenerate wavelengths is difficult in itself. The additional objectives of engineering emission polarization and direction expand the complexity of the task even further.

\begin{figure}[th]
    \centering
    \includegraphics[width=0.9\linewidth]{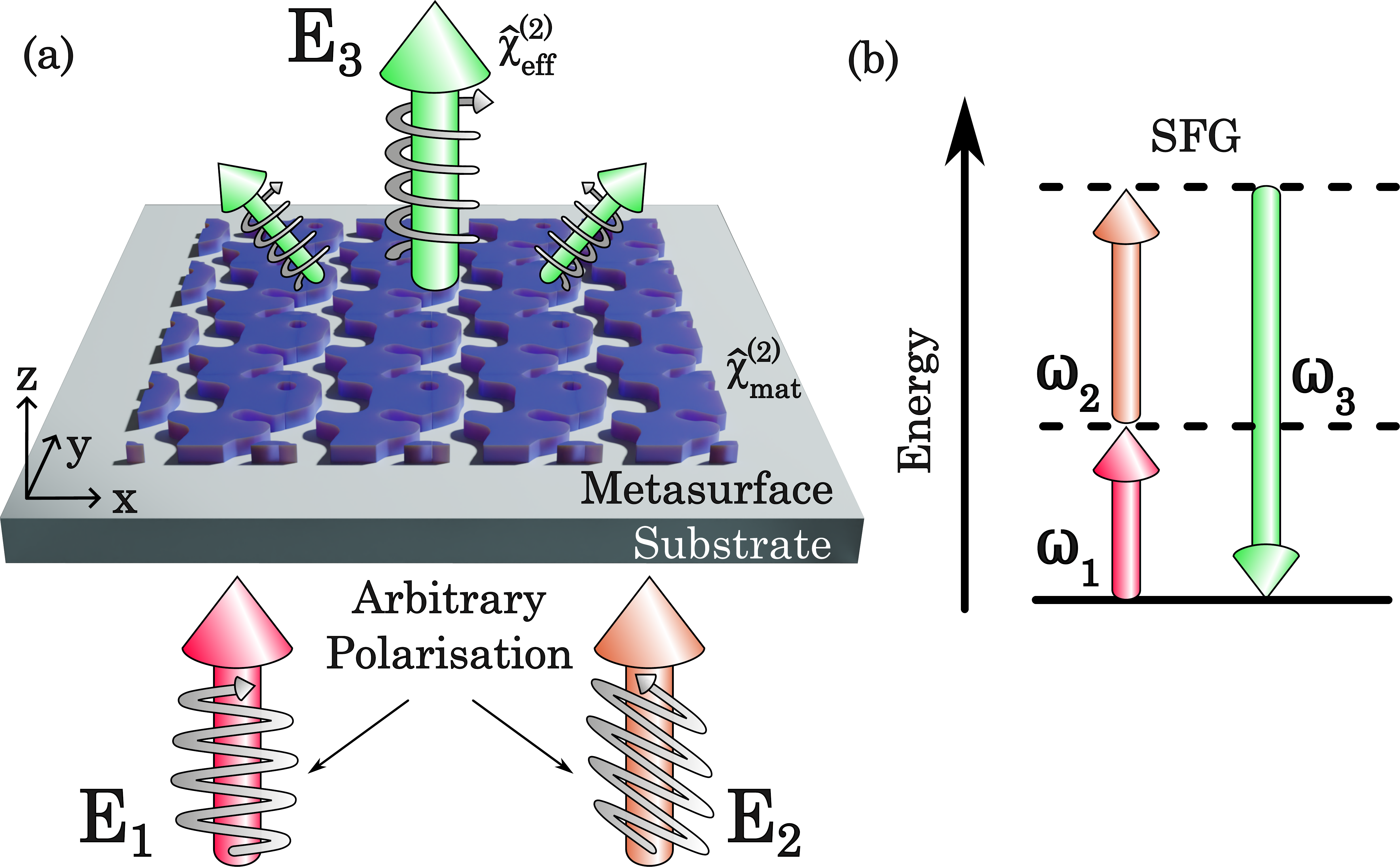}
    \caption{(a) Schematic of the possible processes that can occur through SFG. The input fields $\vb{E}_1$ and $\vb{E}_2$ can have arbitrary polarization states, which then generate the SFG field $\vb{E}_3$. Each diffracted order of the SFG field can have an independent effective $\chi^{(2)}_{\text{eff}}$ associated with it. By optimization of the structure, each of these $\chi^{(2)}_{\text{eff}}$ can be tailored as desired. (b) The energy levels of the SFG process with each field's respective frequencies. }
    \label{fig:Fig1_Schem}
\end{figure}

In this manuscript, we address the aforementioned engineering challenges by developing a novel inverse-design framework for optimizing metasurfaces that enables the simultaneous enhancement of the SFG efficiency over a desired pulse bandwidth, tailoring the polarization transformation matrix, and increasing the emission directionality in a multi-objective manner. Our resulting metasurfaces account for the intrinsic $\hat{\chi}^{(2)}$ of the material and combine it with the optimized structural geometries to form an effective $\hat{\chi}^{(2)}_{\text{eff}}$ of the device [Fig.~\ref{fig:Fig1_Schem}(a)], enabling ultimate control of nonlinear interactions. Thereby,
we broaden the flexibility and increase the benefits of using nonlinear metasurfaces, from wavelength conversion to imaging and other applications.



\section{Methods}
\label{sec:methods}
\subsection{Adjoint Optimization of SFG}

\begin{figure}[!t]
    \centering
    \includegraphics[width=\linewidth]{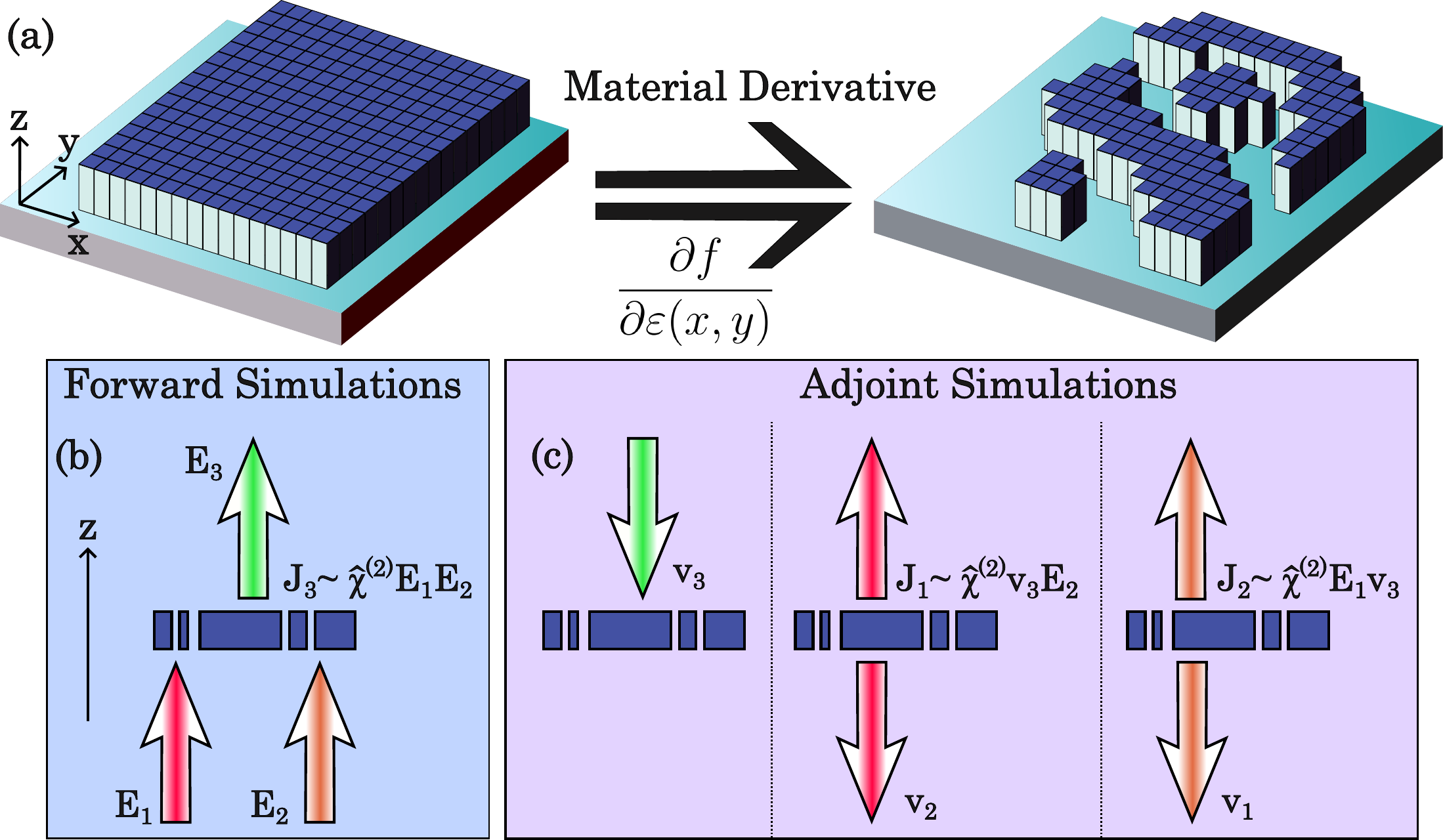}
    \caption{(a) Depiction of topology optimization. The material derivative at each discrete point is calculated through a series of forward and adjoint simulations. (b) Forward simulations of sum-frequency generation, and (c) corresponding adjoint simulations. The forward fields are labelled $\vb{E}_i$, and the adjoint fields are labelled $\vb{v}_j$ for wavelength $i,j$.  }
    \label{fig:Opt_Scheme}
\end{figure}

Beginning from the perspective of classical nonlinear optics, a polarization field is induced in the medium that has a nonlinear relationship with applied electric field \cite{Hillery2009AnOptics, Boyd:2020:NonlinearOptics} 
\begin{equation}
    \vb{P} = \varepsilon_0 \qty[\chi^{(1)}: \vb{E} + \chi^{(2)}: \vb{EE} + \chi^{(3)}: \vb{EEE} + \dots] .
\end{equation}
This polarization then generates a nonlinear current that excites the fields at harmonic frequencies. Our work focuses on the second-order process, so we explicitly write this nonlinear current at the sum-frequency $\omega_3$ as 
\begin{equation}     \label{eq:J3}
    J_{3,i}(\vb{r}) = -i\omega_3\varepsilon_0 \xi(\vb{r}) \sum_{jk}\chi^{(2)}_{ijk} E_{1,j}(\vb{r})E_{2,k}(\vb{r})
\end{equation}
for input fields $\vb{E}_1$ and $\vb{E}_2$ at the frequencies $\omega_1$ and $\omega_2$, respectively [Fig. \ref{fig:Fig1_Schem}(b)]. The implementation of an additional position-dependent parameter $\xi(\vb{r})$ allows the strength of the $\chi^{(2)}$ interaction to be modified simultaneously with the value of the refractive index in the presence or absence of material [Fig.~\ref{fig:Opt_Scheme}(a)]. We assume the non-depleted pump approximation, where the driving fields will not suffer a significant loss of intensity as it propagates due to low enough conversion efficiency. Then the fields $\vb{E}_1$ and $\vb{E}_2$ are found as solutions of linear Maxwell's equations, and then $\vb{E}_3$
at $\lambda_3$ may be obtained by solving the inhomogeneous wave equation with the current from Eq.~(\ref{eq:J3}). 

We aim to maximize the value of an objective function that depends on the generated sum-frequency field, which in turn depends on the medium parameters ($p$) and on the incident fields:
\begin{equation}
    T( \vb{E}_3 ) = T( \vb{E}_3(p, \vb{E}_1, \vb{E}_2) ) . 
\end{equation}
We choose the optimization parameter as $p = \xi({\vb{r}})$ and define the permittivity at each position of the design space as
\begin{equation} \label{eq:eps_p}
    \varepsilon(\vb{r}) = \varepsilon_c + \xi({\vb{r}})(\varepsilon_d-\varepsilon_c), ~0 \leq \xi({\vb{r}}) \leq 1,
\end{equation}
where $\varepsilon_d$ and $\varepsilon_c$ are the permittivity for the patterned and cladding materials, respectively. Thereby, $p = 0$ or $p = 1$ corresponds to the absence or presence of a nonlinear material at a particular spatial location according to Eqs.~(\ref{eq:J3}) \& (\ref{eq:eps_p}), while the values of $0<p<1$ are used during the intermediate stages of the optimization process as we discuss in the following paragraph.


We formulate the adjoint optimization approach to nonlinear metasurfaces following the general principles developed in Refs.~\cite{Jensen:2011-308:LPR, Lalau-Keraly:2013-21693:OE, Niederberger:2014-12971:OE, Molesky:2018-659:NPHOT, Christiansen:2020-4444:OE, Christiansen:2021-496:JOSB}. Previously, nonlinear cavities~\cite{Lin:2016-233:OPT} and metasurfaces~\cite{Sitawarin:2018-B82:PRJ, Mann2023InverseGeneration} were optimized to maximize the total second-harmonic generation. Here, we focus on the SFG process, which includes three fields with different spatial distributions, adding further complexity to the optimisation problem. Furthermore, we perform optimization for SFG radiation over targeted diffraction orders, rather than just the total converted radiation in all directions, as well as the tailored nonlinear polarization transformations.
For this purpose, we aim to computationally efficiently calculate the derivatives $d T / d p$ simultaneously for all spatial points in the nonlinear layer, which then allows for the fast gradient descent optimization of the metasurface patterns as sketched in Fig.~\ref{fig:Opt_Scheme}(a). 
Instead of repeating the forward calculation for material variations at every spatial point [Fig.~\ref{fig:Opt_Scheme}(b)], we identify the adjoint linear problems at the three wavelengths [Fig.~\ref{fig:Opt_Scheme}(c)], which solutions allow the calculation of objective function derivative at all spatial locations through the overlaps between the fields at the three non-degenerate wavelengths and each of their corresponding adjoint fields. Detailed mathematical expressions for the material derivative and its formulation can be found in Supplementary~S1.  In the main manuscript, we summarize the ways in which the necessary adjoint electric fields are obtained. 

For optimization of sum-frequency emission in the outward far-field and on the surface $\bf{\Omega}$, we can define the objective function through the complex amplitude $a$ of a particular wave or mode with a field $({\bf E}_{3,f}, {\bf H}_{3,f})$. This mode can be of any form, including plane waves, Gaussian beams, vortices, and other beam shapes, in combination with any polarization structure. Then, the mode amplitude can be defined as 
\begin{equation} \label{eq:Tlinear}
   a_3  =  \frac{1}{N_3} \iint_\Omega  {\bf n} \cdot \left[ 
        {\bf E_3} \times {\bf H}_{3,b} - {\bf E}_{3,b} \times {\bf H_3}
      \right] d\Omega \,,
\end{equation}
where $\textbf{n}$
is the unit normal vector outward the surface $\Omega$, $({\bf E}_{3,b}, {\bf H}_{3,b})$ is a direction-reversed wave from $({\bf E}_{3,f}, {\bf H}_{3,f})$, and the normalisation coefficient is
$N_3 =  \iint_\Omega  {\bf n} \cdot \left[  {\bf E}_{3,f} \times {\bf H}_{3,b} - {\bf E}_{3,b} \times {\bf H}_{3,f}
      \right] d\Omega$.

We now consider the most common case, in which all materials are reciprocal with symmetric permittivity and permeability, and we only have the electric field-induced electric current source. 
Then, the adjoint field $\vb{v}_{E,3}$ at $\lambda_3$ is a result of linear scattering from the metasurface for a source whose input wave is $({\bf E}_{3,b}/N_3, {\bf H}_{3,b}/N_3)$.

The adjoint fields $\vb{v}_{E,1,2}$ at $\lambda_{1,2}$ are obtained by solving Maxwell's equations with a current source 
\begin{eqnarray} \label{eq:Jv1_SFG1}
      {\bf J}_{v,1}  & = &
        {\bf L}_1^T{\bf v}_{E,3}\,, \\
     \label{eq:Jv2_SFG1}
        {\bf J}_{v,2}  & = &
        {\bf L}_2^T{\bf v}_{E,3}\,, 
\end{eqnarray}
where 
\begin{eqnarray}
    L_{1,ij} =  -i \varepsilon_0 \omega_3 \xi(\vb{r})  \sum_k \chi^{(2)}_{ijk} E_{2,k} \,, \\
    L_{2,ik} =  -i \varepsilon_0 \omega_3 \xi(\vb{r})  \sum_j \chi^{(2)}_{ijk} E_{1,j} \,.
\end{eqnarray}
These currents only exist where $\xi(\vb{r}) \chi_{ijk}^{(2)}$ is non-zero, i.e. in the nonlinear material. We note that these currents have a dependency on $\vb{v}_{E,3}$, and so it must be obtained before solving the aforementioned equations. 

Finally, the objective function gradient is
\begin{equation} \label{eq:dadp_SFG}
\begin{split}
    \frac{d T}{d \xi} = \frac{d a_3}{d \xi} =
       & - i\varepsilon_0\omega_3 \sum_{ijk} \chi^{(2)}_{ijk} E_{1,j}E_{2,k}  \text{v}_{E,3,i} 
        \\ 
       & - i \sum_{q} \omega_{q} {\bf v}_{E,q}^T 
              (\varepsilon_{q,d}-\varepsilon_{q,c}) {\bf E}_{q}         
       ,
\end{split}
\end{equation}
where indices $q=1,2$ are for $\lambda_{1,2}$. 
This equation allows the calculation of derivatives for any functions $T(a_3)$ using a chain rule. For example, for the maximization of SFG light power into a particular mode, we can set $T = \abs{a_3}^2$, and obtain 
\begin{equation} \label{eq:dTdp_SFG}
    \frac{d |a_3|^2}{d \xi} = 
       2\, \text{Re} \left\{ a_3^\ast \frac{d a_3}{d \xi} \right\} .
\end{equation}
In total, a set of six simulations is required to calculate the gradient using the above equations for an arbitrarily large number of spatial positions determined by the computational grid: three forward simulations to calculate ${\bf E}_q$, and three adjoint simulations for ${\bf v}_{E,q}$ at the three non-degenerate wavelengths $\lambda_q$ with $q=1,2,3$, respectively.


\subsection{Optimization for Multiple Polarizations in SFG}


We now discuss the methodology for multi-objective optimization by extending the results in the previous section that were formulated assuming fixed input waves ${\bf E}_1$ and ${\bf E}_2$.
%
Of particular interest is 
to simultaneously tailor the sum-frequency polarization states for multiple combinations of different input polarizations. Then, we can define a figure of merit as
\begin{equation}
    {\cal F}\left( \{a_3^{(m)}({\bf E}_1^{(m)},{\bf E}_2^{(m)})\}_{m=1,\ldots,M} \right) ,
\end{equation}
where $m$ enumerates different input wave combinations and $a_3^{(m)}$ are the sum-frequency amplitudes of the chosen polarization and spatial mode profiles.
Then, the derivative can be found using a chain rule
\begin{equation} \label{eq:chain_rule}
\begin{split}
    \frac{d {\cal F}}{d \xi} = &\sum_{m=1}^M 
    \frac{\partial {\cal F}}{\partial a_3^{(m)}}
    \frac{d a_3^{(m)}({\bf E}_1^{(m)},{\bf E}_2^{(m)})}{d \xi} \\
    &+ 
    \sum_{m=1}^M \frac{\partial {\cal F}}{\partial a_3^{\ast(m)}}
    \frac{d a_3^{\ast(m)}({\bf E}_1^{(m)},{\bf E}_2^{(m)})}{d \xi} ,
\end{split}
\end{equation}
where the derivatives on the right-hand-side are determined using the adjoint formulation in Eq.~(\ref{eq:dadp_SFG}).

In the most general case, the input polarizations for $\lambda_{1,2}$ can each be decomposed into pairs of orthogonal polarization states. Each of $M=4$ combinations of input waves can generate a different nonlinear current. Then, a total of 14 unique simulations are required for the material derivative to fully capture all possible input and output polarization combinations, including 8 forward (2 at $\lambda_{1,2}$ and 4 at $\lambda_{3}$) and 6 adjoint (2 at each wavelength) simulations. 

Furthermore, our method can also optimize diffraction outputs by specifying the adjoint source waves' $k$-vectors. Diffraction into particular orders can be enhanced (suppressed) by increasing (decreasing) the corresponding mode overlap $a_3$, respectively.

\subsection{Numerical implementation}

We iteratively update the function $\xi(\vb{r})$ at all spatial locations inside the quadratically nonlinear material. At each iteration, the material permittivity is updated via gradient descent, or another gradient-based method, at each discretized point in the domain \cite{Molesky:2018-659:NPHOT, Elsawy:2020-1900445:LPR}. For a single-layer metasurface design, we consider the pattern to be independent of the longitudinal coordinate $z$. The optimization concludes when the FOM no longer increases from one iteration to the next or the set maximum number of iterations has been reached. 

In our optimization, we employ several different techniques to ensure that the free-form structures converge to a state that is both physical and realisable. We introduce an increasing binarisation factor as the optimization progresses that eventually forces the final pattern to be either material or air \cite{Ballew2023ConstrainingApplications, Li2022EmpoweringApplications}. The patterns are also periodically blurred with a Gaussian filter that ensures the minimum feature is larger than limitations imposed by fabrication precision \cite{Wang2019RobustMetasurfaces, Panda2020RobustFilters}. Importantly, an artificial absorption coefficient is added onto the material, which will prevent convergence to structures that have arbitrarily large quality factors (Q factor)~\cite{Lin:2016-233:OPT, Liang2013FormulationStates}. This non-radiative decay rate defines the finite bandwidth of the SFG process, which is a practical consideration for future experimental verification and distinguishes our optimisation algorithms from other works purely relying on high-quality factor resonances. This artificial absorption coefficient is later removed for the analysis of the actual metasurfaces.

\section{Results}
\label{sec:results}

\begin{figure*}[!t]
    \centering
    \includegraphics[width=\linewidth]{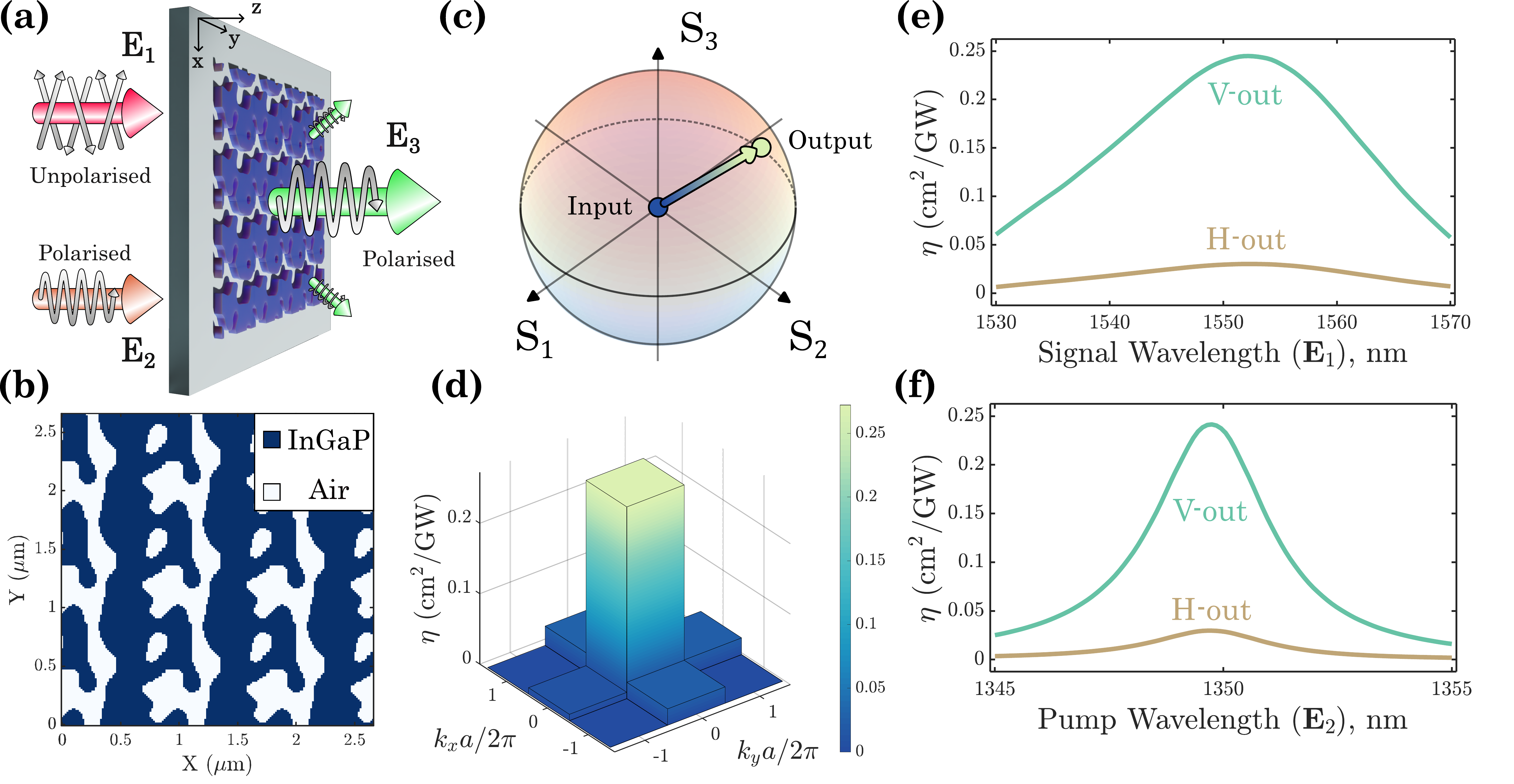}
    \caption{(a) Schematic of an SFG polarizing metasurface. The state of $\vb{E}_2$ is fixed, while the state for $\vb{E}_1$ is unpolarized. The SFG light from the optimized metasurface is polarized in the $\ket{V}$, with the higher diffraction orders also being suppressed. (b) optimized pattern of the metasurface. (c) Poincar\'{e} sphere representation of all input states of $\vb{E}_1$ being transformed into the same SFG output ($S_1$, corresponding to $\ket{H}$) state. (d) SFG efficiency from an unpolarized source for different diffraction orders. (e) Signal wavelengths sweep of SFG efficiency spectra for a pump wavelength fixed at \SI{1350}{\nano\metre}, and (f) pump wavelengths sweep of SFG efficiency spectra for a signal wavelength fixed at \SI{1550}{\nano\metre}.  }
    \label{fig:Pol_Spec}
\end{figure*}

In general, we can optimize for any combination of polarizations and elements in the effective $\chi^{(2)}$ tensor, as discussed above. However, in this work, we tackle a simpler problem that nevertheless still highlights the strength of our optimization scheme. Specifically, we consider a signal having any transverse polarization while the pump has a fixed polarization. Such functionality may be beneficial for the operation of upconversion infrared imaging, where the image can have any arbitrary polarization and the pump is a light source of fixed polarization. For our examples, we have a signal wavelength $\lambda_1 = \SI{1550}{\nano\metre}$ and a pump wavelength $\lambda_2 = \SI{1350}{\nano\metre}$ (resulting in $\lambda_3=721.6$~nm). The transformation of the signal state into the SFG state can be expressed in the form 
\begin{equation}
    \vb{E}_{3} = \mathbf{M}\qty(\vb{E}_2,\chi^{(2)},\varepsilon)\cdot\vb{E}_1.
    \label{eq:nl_Jones}
\end{equation}
The fields $\vb{E}_q$ represent the orthogonal complex amplitudes $\qty(E_q^x, E_q^y)$ in the transverse components basis. We can make a connection of Eq.~(\ref{eq:nl_Jones}) to the Jones vectors in polarization optics in the following way. Let $\mathbf{M}$ represent the nonlinear scattering matrix analogous to the Jones matrix in linear materials. Individual elements of $\mathbf{M}$ can be interpreted as complex scattering amplitudes. For an unpatterned film, $\mathbf{M}$ is constrained by the crystal orientation and is set by the elements of $\chi^{(2)}$ and the input state of $\vb{E}_2$. One such case is provided in Supplementary~S3.1 for unpatterned nonlinear film. Then $\vb{E}_1$ and $\vb{E}_3$ can be interpreted as the input and output Jones vectors of the system, respectively. Now, with the ability to pattern structures into the nonlinear film, the permittivity $\varepsilon$ is no longer uniform across the domain but instead can be engineered to achieve the desired transformation of $\mathbf{M}$. 

In all the examples, the nonlinear material is indium gallium phosphide (InGaP) that is of (100) crystalline orientation and \SI{300}{\nano\metre} thick. The film is resting on the fused silica substrate, with a $\SI{900}{\nano\metre} \times \SI{900}{\nano\metre}$ unit cell. See Supplementary~S2 for the details regarding InGaP parameters. The electromagnetic simulations are performed using the commercial COMSOL Multiphysics software package suite. Each simulation takes approximately 100 seconds when using a discretization of \SI{40}{\nano\metre}. The resulting fields are exported to the MATLAB programming language, where we implement our inverse-design optimization.

\subsection{Polarizing Nonlinear Metasurface}
\label{subsec:pol_met}

At the SFG wavelength, multiple diffraction orders exist due to the periodicity of the metasurface. In 
infrared imaging applications, the higher diffraction orders should be ideally suppressed so that the majority of the SFG light is propagating in the zeroth order. In previous works \cite{Camacho-Morales2021InfraredMetasurfaces,Camacho-Morales2022Sum-FrequencyModes}, it has been a challenge to suppress these higher-order propagating modes. Simultaneously, with our method, we can also optimize diffraction outputs by specifying the adjoint source waves' $k$-vectors. 
This is a significant advance compared to previous inverse-design approaches for quadratically nonlinear metasurfaces \cite{Sitawarin:2018-B82:PRJ, Mann2023InverseGeneration} where only total second-harmonic conversion but not directionality could be optimized.

We denote the zeroth order of the scattering matrix as $\mathbf{M}_0$, whose elements can be optimized individually. In the first example, the signal is an unpolarized state while the pump at $\lambda_2$ is polarized in the $\ket{V}$ state. We intend for the SFG zeroth order diffraction to also be polarized in the $\ket{V}$ state [Fig.~\ref{fig:Pol_Spec}(a)]. In this scheme, the metasurface can be considered to be a nonlinear polarizer by taking an unpolarized input and polarizing it into the $\ket{V}$ state at the SFG output. The figure of merit is formulated as 
$
    {\cal F} = \frac{1}{2}\qty(\abs{\bra{V_3}\mathbf{M}_0\ket{H_1}}^2 + \abs{\bra{V_3}\mathbf{M}_0 \ket{V_1}}^2)
$, 
and its derivative is calculated with Eq.~(\ref{eq:chain_rule}). 
We note that unpolarized light can be decomposed into equal powers of any pair of orthogonal states, which are the $\ket{V}$ and $\ket{H}$ pair in our modelling. In this notation, the states $\ket{\psi_1}$ denote the input signal state, while $\bra{\phi_3}$ denote the outgoing SFG state. 

The optimization converges to a design that is highly non-trivial, as shown in Fig.~\ref{fig:Pol_Spec}(b). We calculate the expected transformation of input unpolarized light and find that it is almost fully converted into the $\ket{V}$ state at the SFG wavelength, as depicted on the Poincar\'{e} sphere in Fig.~\ref{fig:Pol_Spec}(c). The numerical values of $\mathbf{M}_0$ matrix elements can be found in Supplementary~S3.2, where further analysis is also provided. In Fig.~\ref{fig:Pol_Spec}(d), we show the predicted SFG conversion efficiency into all the propagating diffraction orders, of which the zeroth order comprises approximately 80\% of the total SFG light (see Supplementary~S4.1 for precise values). Therefore, the optimized metasurface directs the vast majority of SFG light perpendicular to the surface, which greatly benefits imaging applications. 

We now determine the frequency bandwidth by fixing the wavelength of the pump at $\lambda_2 = \SI{1350}{\nano\metre}$ and calculating the SFG efficiency as an unpolarized signal wavelength is swept around the operating wavelength of $\lambda_1 = \SI{1550}{\nano\metre}$ [Fig.~\ref{fig:Pol_Spec}(e)]. The transmission into the desired $\ket{V}$ state is significantly larger than the $\ket{H}$ state and reaches a maximum SFG efficiency of \SI{0.25}{\centi\metre\squared\per\giga\watt}, which is more than 3 orders of magnitude larger than an unpatterned film of the same nonlinear material and thickness. This performance is echoed when we fix the signal $\lambda_1 = \SI{1550}{\nano\metre}$ and sweep the pump wavelengths instead [Fig.~\ref{fig:Pol_Spec}(f)]. The full-width at half maximum (FWHM) conversion  efficiency for the signal is approximately \SI{20}{\nano\metre}, while it is approximately \SI{3}{\nano\metre} for the pump, representing a reasonably large operating bandwidth suitable for efficient conversion with short optical pulses.
From these plots, it is evident that there are resonances present within the metasurfaces at the input wavelengths $\lambda_1$ and $\lambda_2$ that greatly enhance the SFG process. We provide the field distributions and further analysis in 
the Supplementary~S5.  

\subsection{Polarization Independent Nonlinear Metasurface}
\label{subsec:pol_ind_met}

\begin{figure*}[!t]
    \centering
    \includegraphics[width=\linewidth]{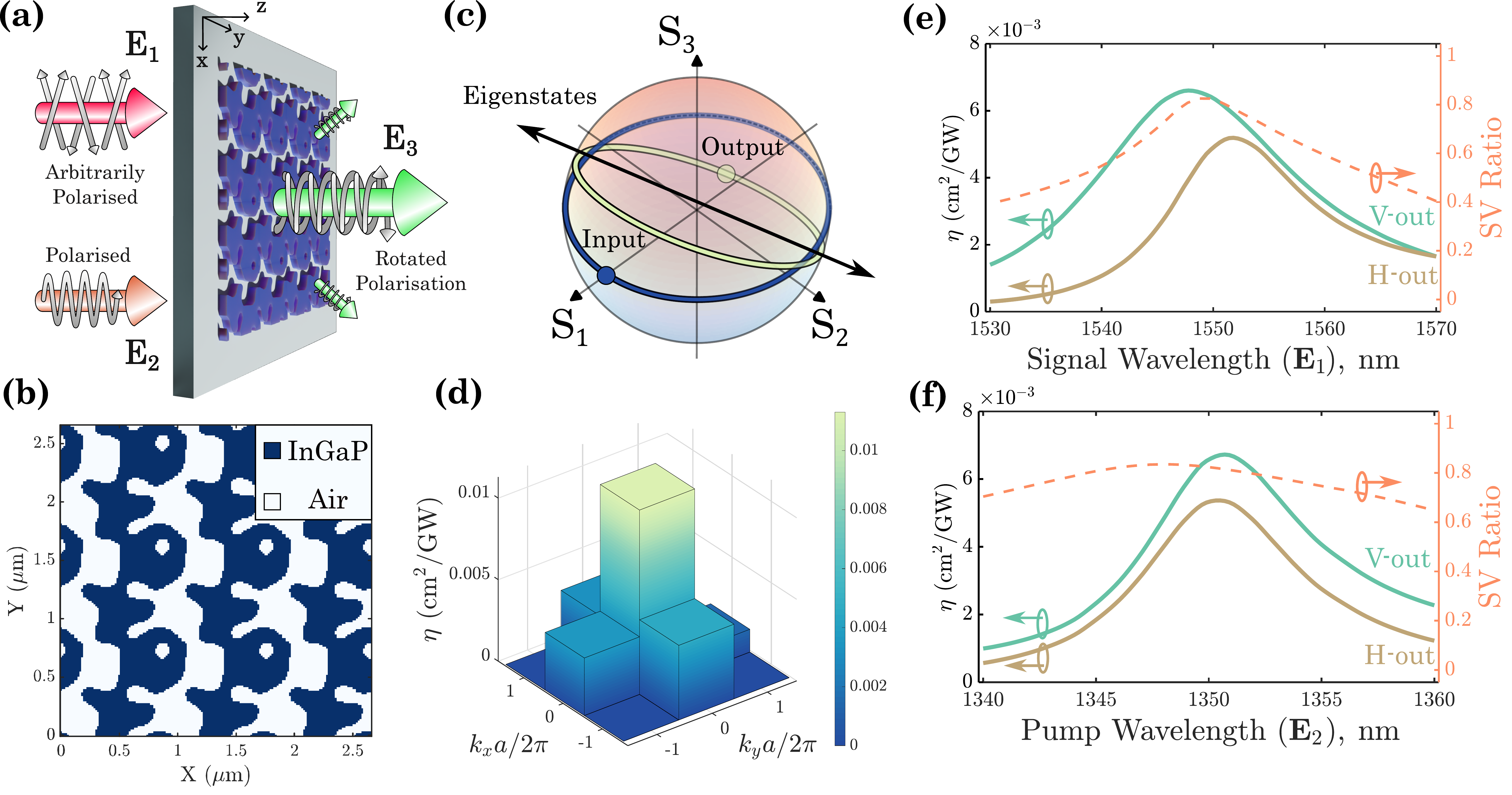}
    \caption{(a) Schematic of an SFG waveplate metasurface. The state of $\vb{E}_2$ is fixed, while the state for $\vb{E}_1$ is unpolarized. The SFG light from the optimized metasurface is rotated in the $\ket{V}$, with the higher diffraction orders also being suppressed. (b) optimized pattern of the metasurface. (c)~Poincar\'{e} sphere representation of all input states of $\vb{E}_1$ being transformed into different SFG output states while preserving their orthogonality. (d)~SFG efficiency from an unpolarized source for different diffraction orders. (e)~Signal wavelengths sweep of SFG efficiency spectra for a pump wavelength fixed at \SI{1350}{\nano\metre}, and (f)~pump wavelengths sweep of SFG efficiency spectra for a signal wavelength fixed at \SI{1550}{\nano\metre}. The dashed orange curves show the ratio of singular values (SV) for each plot at their respective wavelengths. }
    \label{fig:MinS_Spec}
\end{figure*}

In this example, we focus on a metasurface whose SFG enhancement is independent of the polarization of $\lambda_1$. A metasurface that has this property can enhance SFG conversion efficiency for all input polarization states equally. This is particularly useful for imaging, where the source is typically unpolarized or partially polarized. Such an upconverted image will preserve the relative amplitudes of the original image, even if there are spatial variations in the polarization. This is in contrast to previous demonstrations of SFG imaging, where the metasurfaces typically rely on polarization-sensitive resonant modes of bound states in the continuum \cite{Camacho-Morales:2021-36002:ADP}. Therefore, for this case, the target $\mathbf{M}_0$ matrix is close to unitary after scaling, or its singular values are close to equality. The FOM for this case is 
\begin{equation}
    {\cal F} = s_2 \,,
\end{equation}
where $s_1$ and $s_2$ are the ordered singular values of $\mathbf{M}_0$, and for this demonstration, the pump polarization at $\lambda_2$ is fixed at $\ket{V}$. By maximising $s_2$, which is always defined as the smaller of the two singular values, we are effectively increasing the SFG enhancement of the worst-performing polarization input state. For a unitary matrix, the ratio of the singular values must be $s_2/s_1 = 1$.

The optimized metasurface is again a non-trivial free-form pattern, as shown in Fig.~\ref{fig:MinS_Spec}(b). For this metasurface, we then calculate the transformation of various input states at the signal wavelength and plot them on a Poincar\'{e} sphere [Fig.~\ref{fig:MinS_Spec}(c)]. We see that the metasurface imparts a rotation of the eigenstates to the input states during the SFG process. We provide further analysis of the matrix $\mathbf{M}_0$ in Supplementary~S3.3. Importantly, the eigenstates are nearly orthogonal, which indicates that the transformation is indeed near unitary after scaling. The SFG is primarily channelled into the zeroth diffraction order for an unpolarized signal [Fig.~\ref{fig:MinS_Spec}(d)], with reduced light leakage into higher order diffraction modes (see Supplementary~S4.2 for numerical values). 

We calculate the SFG efficiency for a range of unpolarized input signal wavelengths and pump wavelength fixed at $\lambda_2 = \SI{1350}{\nano\metre}$ and find a maximum in average efficiency of \SI{6d-3}{\centi\metre\squared\per\giga\watt} [Fig.~\ref{fig:MinS_Spec}(e)]. Again in this example, the FWHM conversion efficiency for the signal is approximately \SI{20}{\nano\metre}, while it is approximately for \SI{10}{\nano\metre} for the pump. The SFG efficiency for two orthogonal polarizations of $\ket{H}$ and $\ket{V}$ are almost equal. On the right axis, we show the ratio of singular values that reaches a maximum of 0.8, reasonably close to the ratio of 1 for a truly unitary transformation. This analysis is repeated for a fixed signal ($\lambda_1$ = \SI{1550}{\nano\metre}) and varying pump wavelengths [Fig.~\ref{fig:MinS_Spec}(f)], with the maximum efficiency peaking at $\lambda_2$ = \SI{1350}{\nano\metre}, as expected. Therefore, the transformation of input light into SFG output from the metasurface can be considered to be nearly polarization-independent. The preservation of polarization information leads to the ability to perform upconversion polarimetric imaging \cite{Zhu2022NonlinearUpconversion} with greater efficiency than previously reported.

\subsection{Structure of Resonances in the Optimised Metasurfaces}
\label{sec: discussio}

Finally, we perform linear simulations to inspect the resonances that we expect to be present in the metasurfaces at $\lambda_1,\lambda_2,\lambda_3$ (Supplementary~S5.1). Because the two metasurfaces presented in this work in Sec.~\ref{subsec:pol_met} and~\ref{subsec:pol_ind_met} above are optimized for different functionalities, they also have different resonant characteristics. For the nonlinear polarizer metasurface, only $\ket{H}$ produces a resonance at $\lambda_1$, while for the polarization-independent nonlinear metasurface, resonances appear for both $\ket{H}$ and $\ket{V}$ close to $\lambda_1$. As expected for both nonlinear metasurfaces, only $\ket{V}$ produces a resonance at the pump wavelength because the polarization at $\lambda_2$ was fixed. We note that the $Q$-factors 
are on the order of 70 to 300 for the different wavelengths, which allows for a reasonably broad imaging bandwidth and for SFG with short optical pulses. We also provide field enhancement distributions for the metasurfaces at all the interacting wavelengths 
in Supplementary~S5.2.


\section{Conclusions}
\label{sec:conclusion}
We have developed a novel method of multi-objective optimization of nonlinear frequency mixing processes in metasurfaces. Our method allows for the simultaneous control of polarizations and directionality and maximizes efficiency across a target bandwidth for SFG processes, which is beyond what is possible with conventional design strategies. We present a computationally efficient implementation based on adjoint formulation and demonstrate two essential examples of metasurfaces that enhance SFG either for one signal input polarization or for all input polarizations.
In both cases, the SFG is emitted in the forward direction while the higher diffraction orders are suppressed.

This method can be naturally extended to optimize metasurfaces that explore other nonlinear phenomena such as four-wave mixing, spontaneous parametric down-conversion, and high harmonic generation with greater efficiency. In the future, nonlinear metasurfaces that exhibit more complicated characteristics will be enabled by sophisticated optimization algorithms, widening the spectrum of potential functionalities. 

\begin{funding}
    This work was supported by the Australian Research Council (CE200100010).
\end{funding}

\begin{authorcontributions}
All authors have accepted responsibility for the entire content of this manuscript and approved its submission.

\end{authorcontributions}

\begin{conflictofinterest}
The authors state no conflict of interest.
\end{conflictofinterest}

\begin{dataavailabilitystatement}
The datasets generated during and/or analyzed during the current study are available from the corresponding author upon reasonable request.
\end{dataavailabilitystatement}

\printbibliography

\end{document}